# An Anomaly in the Angular Distribution of Quasar Magnitudes: Evidence for a Bubble Universe with a Mass ~$10^{21}$ M$_\odot$

Michael J. Longo[1]

Department of Physics, University of Michigan, Ann Arbor, MI 48109, USA

**Abstract**

Quasars provide our farthest-reaching view of the Universe. The Sloan Survey now contains over 100,000 quasar candidates. A careful look at the angular distribution of quasar magnitudes shows a surprising intensity enhancement with a "bulls eye" pattern toward ($\alpha$, $\delta$) ~ (195°, 0°) for all wavelengths from UV through infrared. The angular pattern and size of the enhancement is very similar for all wavelengths, which is inconsistent with a Doppler shift due to a large peculiar velocity toward that direction. The enhancement is also too large to explain as a systematic error in the quasar magnitudes. The general features of the anomaly can be explained by the gravitational lensing of a massive bubble with $M_{lens}$ ~ $10^{21}$ M$_\odot$ and a lens radius ~350 Mpc; its effects extend over an angular range ~ ±30° on the sky. It is remarkable that the presence of such a massive bubble universe can explain not only the anomalies in the angular distribution of quasar intensities, but also anomalies in the distribution of luminous red galaxies, anomalies in the CMB, and bulk flow discrepancies, all of which appear in roughly the same direction.

## 1. Introduction

Quasars, by virtue of their enormous luminosities, offer our farthest-reaching picture of the Universe. The Cosmological Principle requires their distributions be homogeneous when averaged over sufficiently large volumes of space. By now, the Sloan Digital Sky Survey (SDSS) [1] has found over 100,000 quasar candidates [2] out to redshifts $z$ beyond $z = 5$. These have well measured redshifts based on spectral lines and relative magnitudes for 5 contiguous wavelength bands ranging from 400 nm to 1000 nm [3], which correspond to near ultraviolet *U*, green *G*, red *R*, infrared *I*, and far-infrared *Z*. The filter magnitudes can be thought of as a measure of the apparent brightness of the quasar in each color band. The *U*-band filter magnitudes have a typical accuracy of 0.03 mag for each quasar and half that for the other bands. Quasar candidates are selected via their nonstellar colors in *UGRIZ* broadband photometry [4]. The selection is based upon measurements that have been corrected for Galactic extinction using the maps of Finkbeiner, Davis, and Schlegel [5].

---

[1] email: mlongo@umich.edu

The distribution of the quasar sample in redshift is highly biased because the selection criteria are explicitly $z$ dependent. There is also the natural falloff at large $z$ as the quasars become fainter. However, it is important to note that the selection criteria depend explicitly only on the *UGRIZ* magnitudes, and not on the angular coordinates, right ascension $\alpha$ and declination $\delta$. The question of statistical completeness of the sample is discussed in detail by Richards *et al* [4]. These issues affect only the *z*-dependence, and are largely confined to $z > 2.2$, which were not used in this study. The filter magnitudes have been corrected for atmospheric effects, and the magnitudes used in this study were corrected for Galactic extinction [5]. Thus, a systematic dependence on ($\alpha, \delta$) of the measured magnitudes is not expected. The technique used here is therefore to find the *z* dependence of the magnitudes averaged over the entire sample. Then the average magnitude for a given *z* is subtracted from that for each quasar to look for a possible systematic dependence on ($\alpha, \delta, z$). As we shall discuss below, there is a significant enhancement ~0.25 magnitude brighter in all the filter magnitudes in a cone ~±30° wide that is centered at $\alpha$ ~195° and $\delta \sim 0°$ for the redshifts $z > 0.5$.

## 2. Analysis

For this analysis, all of the quasar candidates identified as "QSO" from the SDSS Quasar Catalog V, DR7 [2] were used. Since the density of quasar candidates falls off rapidly beyond $z$~2.2, it was restricted to $z < 2.2$, which corresponds to a distance ~6600 $h^{-1}$ Mpc. Redshifts less than 0.2 were also excluded because the magnitudes were varying rapidly with *z* due to the selection criteria. This left approximately 82,500 quasars. Of these, 75,300 were in the right ascension range 110°< $\alpha$ <255° and declinations –10°< $\delta$ <70°. The remaining 7,200 were in –50°< $\alpha$ <65°, mostly in 3 narrow declination bands near –8°, 0°, and 15°.

Files with the ($\alpha, \delta$) coordinates, redshift, redshift errors, and *UGRIZ* magnitudes corrected for Galactic dust extinction were downloaded from the SDSS CASJOBS website. Much of the subsequent discussion refers to the *U* magnitudes, though the other filter magnitudes showed a similar behavior. Figure 1 shows the smoothed *U* and *Z* filter magnitudes *vs*. redshift for all the quasars in the sample. An estimated absolute *U* magnitude is also shown.

For each quasar the deviation of the $U$ magnitude from the average $U$ magnitude for that redshift was calculated as

$$\Delta U = U(\alpha, \delta, z) - \bar{U}_{int}(z) \qquad (1)$$

where $\bar{U}_{int}(z)$ is the average $U$ for that redshift, as interpolated from the curve in Fig.1. As defined, a negative $\Delta U$ corresponds to the quasar being <u>brighter</u> than average at that redshift.

The sample was then divided into slices in declination. Polar plots of declination vs. $z$ for two of the $\delta$ slices are shown in Fig. 2. The points give the ($\alpha$, $z$) position of each quasar. Their colors indicate $\Delta U$ for that quasar. The colors range from violet through blue, green, orange, and red, with violet representing the largest magnitude enhancement (most negative $\Delta U$), red the largest deficit, and green for $U$ approximately equal to the average. Magnitude contours are also shown in Fig. 2. These were generated by binning the $\Delta U$ for each $\delta$ slice into 16 bins in $\alpha$ and 8 bins in $z$ for a total of 128 bins, with typically ~100 entries per bin. This array was then smoothed and contours plotted. The contours range from $\Delta U = -0.2$ (violet), $-0.1$ (blue), $0.0$ (black), $0.1$ (orange), and $0.2$ (red). The contours and points in the hemisphere toward $\alpha = 0°$ correspond to the entire $\delta$ range there, because of the limited range of declinations in the sample in that hemisphere. Thus they are the same in both $\delta$ slices.

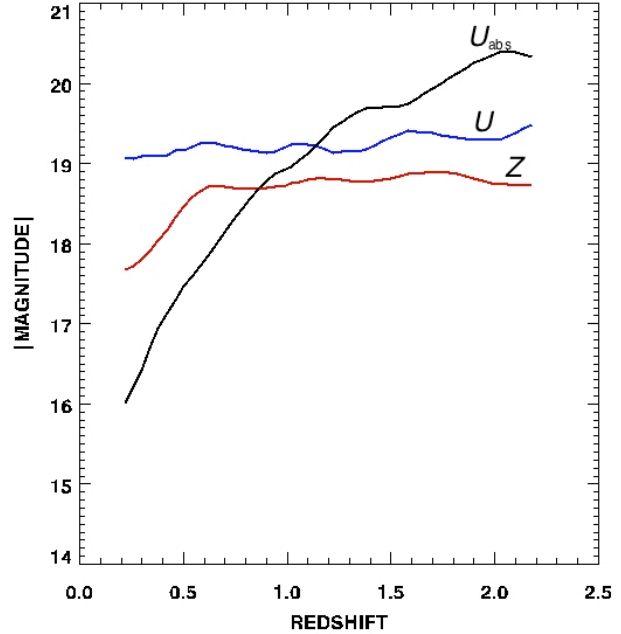

Figure 1 – Smoothed $U$ and $Z$ magnitudes for the entire sample vs. redshift. The black curve shows the approximate absolute $U$ magnitude. Its sign has been dropped for convenience.

A systematic magnitude enhancement toward $\alpha = 180°$ is seen for the $10° < \delta < 20°$ slice. There is no sign of it in the other hemisphere. Six $\delta$ slices in the left hemisphere were studied in $10°$ increments in $\delta$ from $0°$ to $60°$. These are statistically independent and all show similar systematic magnitude shifts, which appear to be greatest in the $0° < \delta < 10°$ and $10° < \delta < 20°$ slices, gradually decreasing toward $\delta = 60°$.

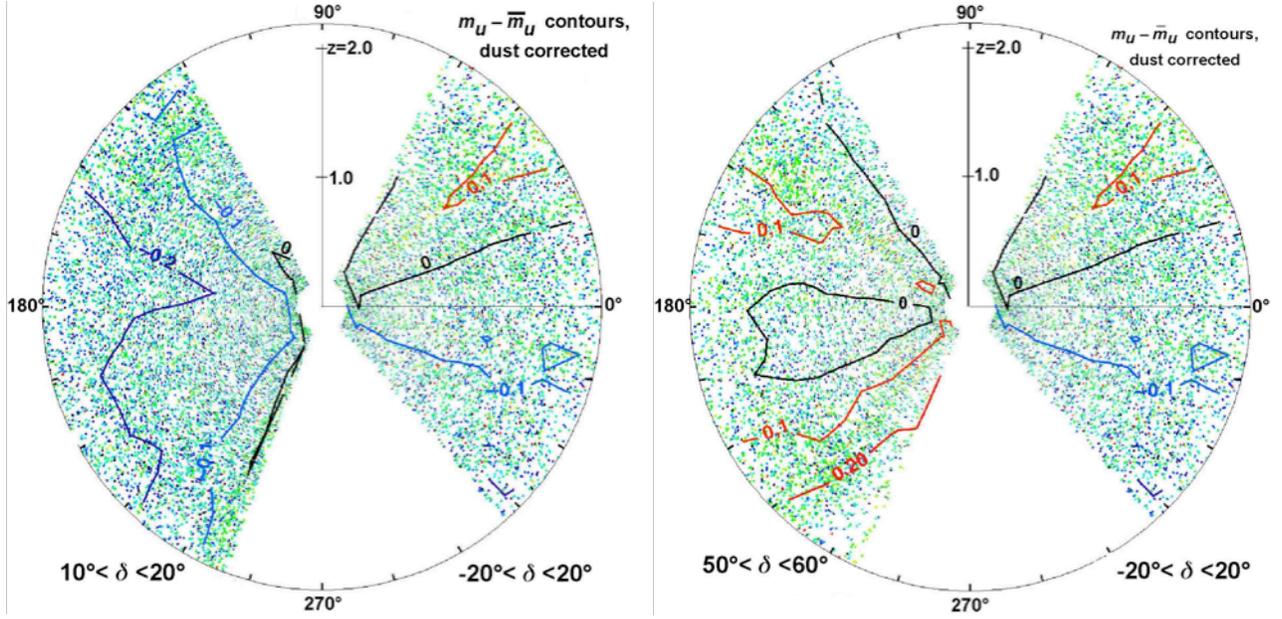

Figure 2 – Polar plots of quasar UV band magnitude shift *vs*. redshift and right ascension for two declination slices. The overall $z$ dependence has been removed. The contours, described in the text, range from $\Delta U = -0.2$ (violet), $-0.1$ (blue), $0.0$ (black), $0.1$ (orange), and $0.2$ (red). The dots indicating the quasar positions have a similar color correspondence, ranging from violet-blue-green-orange-red. The gaps toward 90° and 270° are due to the obscuration by the Galactic disk. The hemisphere toward 0° is the same in both plots since the SDSS coverage there is limited to 3 small $\delta$ ranges between $-15°$ and $15°$. For the hemisphere toward 180° the $\delta$ ranges are 10° to 20° for the left plot and 50° to 60° for the right plot. The enhancement of the luminosities toward 195° for small $\delta$ is apparent in the left plot.

A more quantitative appreciation of the magnitude enhancement toward $\alpha = 195°$ can be gotten by looking at the variation of $\langle \Delta U \rangle$, the average $\Delta U$ for each bin in the 128 $\alpha, z$ bins described previously. This is shown in Fig. 3 for the slices $0° < \delta < 10°$ and $50° < \delta < 60°$. The 8 sets of curves show the variation of $\langle \Delta U \rangle$ with $\alpha$ for each of the 8 $z$ bins. The red (dash-dot) curves on the right side are for the right hemisphere in Fig. 2 translated by 180° to allow an easy comparison. The error bars are calculated from the standard deviation of the $\Delta U$ for the entire quasar sample divided by the square root of the number of entries in the bin. For $0° < \delta < 10°$, all of the $z$ ranges above $z \approx 0.5$ show a dip of about 0.25 mag in $\langle \Delta U \rangle$ for $150° \lesssim \alpha \lesssim 240°$. This dip is much less apparent for the $50° < \delta < 60°$ plots, and is not present in the dash-dot curves for the other hemisphere. The angular correlations, as well as the dips in Fig. 3, disappear if the $\alpha$'s and $\delta$'s are scrambled among the quasars. Note that all of these curves are statistically independent. The other 4 $\delta$ slices, which are also statistically independent, appear to show a similar behavior with a larger dip at smaller declinations. Thus the statistical significance of the effect is overwhelming.

The angular dependence of the effect can be better seen by looking at the $(\alpha, \delta)$ dependence within a spherical shell containing a small range of redshifts. A plot of $\langle \Delta G \rangle$ vs. declination and right ascension for a spherical shell with $1.53 < z < 2.20$ is shown in Fig. 4, where $\langle \Delta G \rangle$ is defined in analogy with $\langle \Delta U \rangle$ for the green filter magnitude. As before, the overall $z$ dependence has been removed, and the contours are based on the smoothed $\langle \Delta G \rangle$, in this case for 16 $\alpha$ and 10 $\delta$ bins. The contours range from $\langle \Delta G \rangle = -0.15$ (violet), $-0.10$ (blue), $-0.05$ (black), 0.0 (orange), to 0.05 (red). The "hotspot" (blue-violet) toward (195°, 5°) is apparent. It covers an angular range ~80° in $\alpha$, and at least 35° in $\delta$. The $\Delta G$'s for the entire sample within this shell must sum to 0, so that the hotspot appears to be surrounded by a colder (less bright) region. In other words, we are only looking at $\langle \Delta G \rangle$ differences over the range of $(\alpha, \delta)$ covered by SDSS. Because the hotspot includes ~40% of the total sample, its contrast with the surroundings is reduced somewhat by the $z$ averaging procedure.

Plots of the $(\alpha, \delta)$ dependence for the $z$ ranges $0.50 < z < 1.0$ and $1.00 < z < 1.53$, which are statistically independent, show a very similar behavior. The angular range and magnitude range of the hotspot is nearly the same for all $z \gtrsim 0.5$; for smaller $z$ it appears to diminish in size and magnitude range. Again, the angular correlations disappear if the $\alpha$'s and $\delta$'s are scrambled among the quasars. The declination range of the hotspot is truncated by the lack of coverage of SDSS below $\delta \sim -5°$, and it probably is centered at a declination below $-5°$.

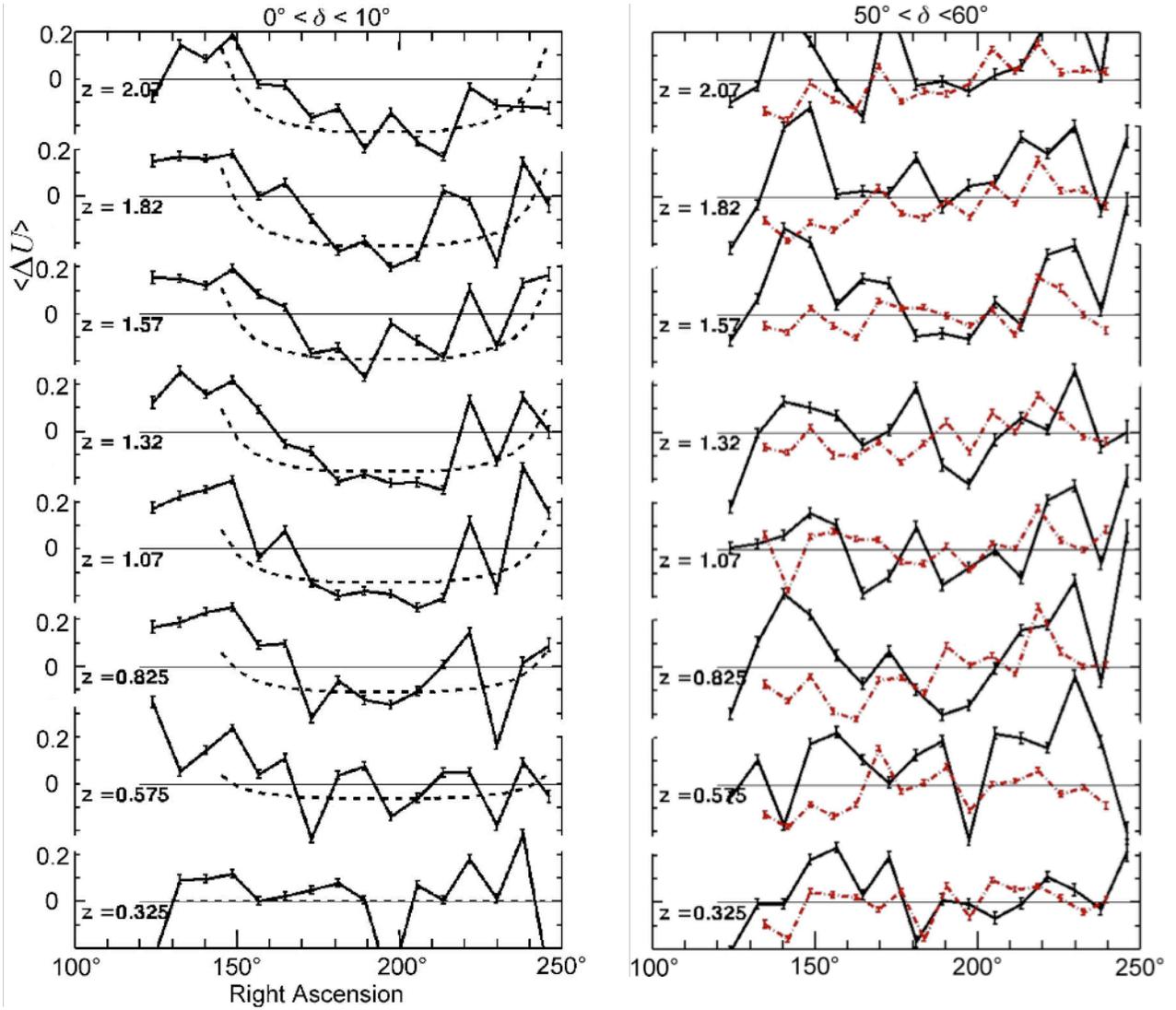

Figure 3 – The variation of $\langle \Delta U \rangle$ with $\alpha$ for each of the $z$ bins and two of the $\delta$ slices. Negative $\langle \Delta U \rangle$ correspond to a higher brightness. The dashed curves in the left plot show the variation expected in the bubble model described in Section 3. The red (dash-dot) curves on the right plot are for the right hemisphere in Fig. 2 translated by 180° to allow an easy comparison.

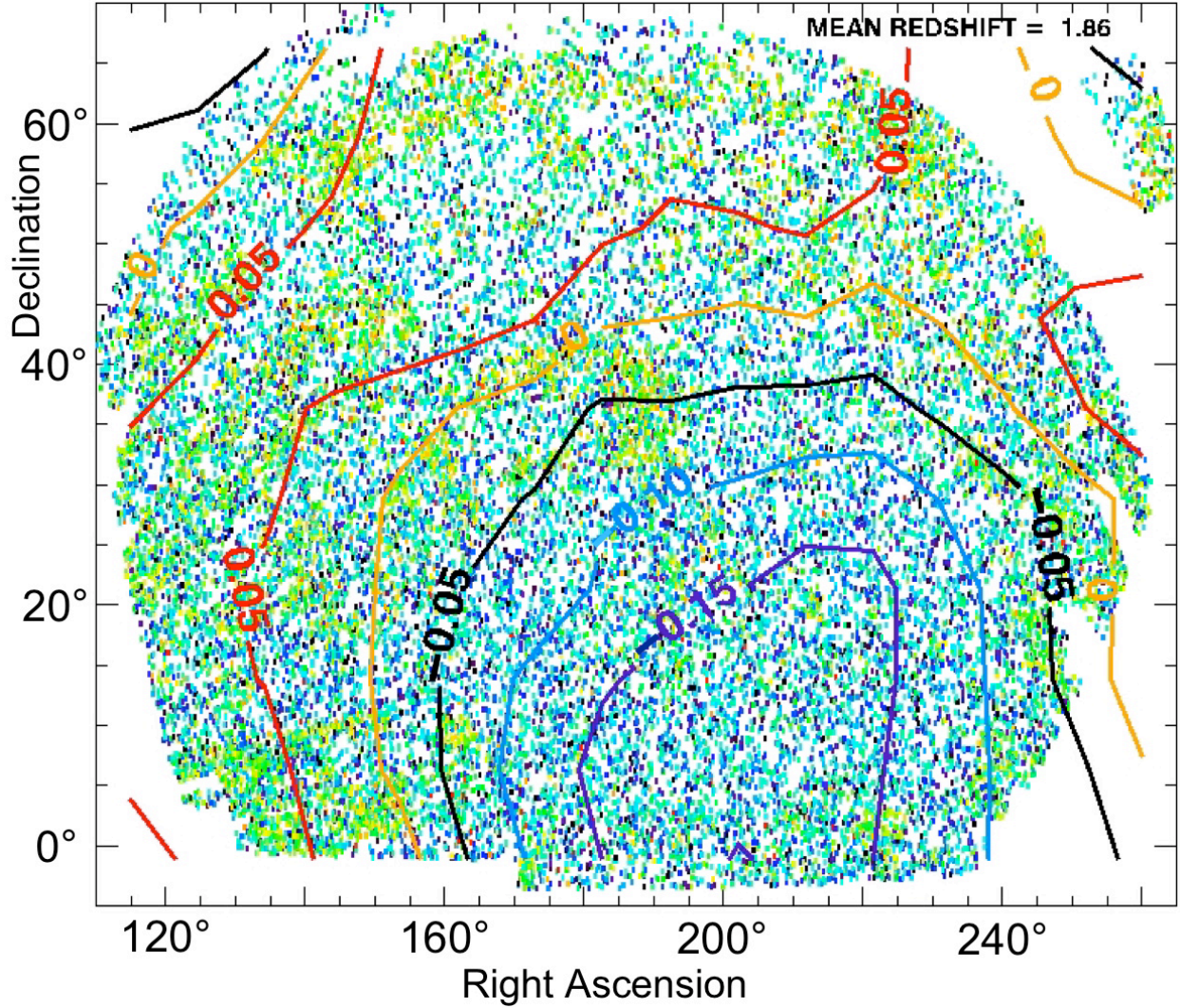

Figure 4 – Plot of green magnitude shift *vs.* declination and right ascension for a spherical shell at a radius 1.53 < *z* < 2.20. The contours range from –0.15 (violet), –0.10 (blue), –0.05 (black), 0.0 (orange), to 0.05 (red). The dots indicating the quasar positions have a similar color correspondence, ranging from violet-blue-green-orange-red. The "hotspot" (blue-violet) toward (195°, 0°) is apparent. The Δ*G*'s for the entire sample within this shell must sum to 0, so that the hotspot appears to be surrounded by a dimmer (redder contours) region. The same "bulls eye" pattern in the angular distribution of the quasar magnitudes appears for all redshift ranges above *z*~0.5 and for all 5 filter bands.

## 3. Gravitational Lens Model

In this section we discuss a simple model for the intensity enhancement produced by a gravitational lens. Figure 5 shows the geometry for a single disk lens. Q is the quasar source at an angular diameter distance $D_{LS}$ from the lens with a total mass $M_{lens}$. The lens is assumed to be a uniform disk of radius R. A ray goes off from the source at an angle $\theta$ and impacts the disk at an

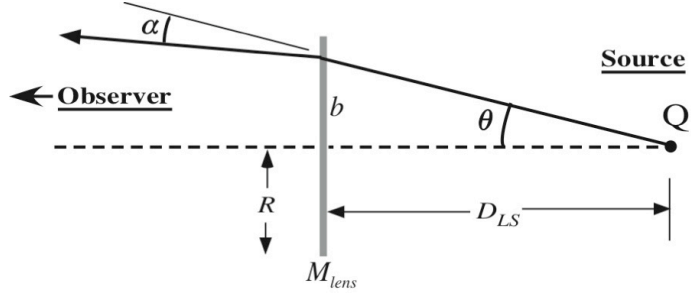

Figure 5 – Geometry for gravitational lensing modeled as a thin disk of mass $M_{lens}$

impact parameter $b$. The deflection angle $\alpha$, is essentially the same as for a point mass:

$$\alpha \cong \frac{4GM(b)}{c^2 b} \approx \frac{4GM(b)}{c^2 \theta D_{LS}} \tag{2}$$

where G is the gravitational constant, $M(b)$ is the mass within the radius $b$, and we use the small angle approximation, $b \approx \theta D_{LS}$. The intensity enhancement or amplification comes about because the light from the source within a small angle $\theta$ is focused into a smaller cone $(\theta - \alpha)$. This compression only occurs in the radial direction so the amplification factor is $\Delta\alpha/\Delta\theta$. This gives

$$\left|\frac{\Delta\alpha}{\Delta\theta}\right| \approx \frac{4GM(b)}{c^2 \theta^2 D_{LS}} \approx \frac{4GM(b)D_{LS}}{c^2 b^2} \tag{3}$$

For a uniform disk, $M(b)=b^2 M_{lens}/R^2$, so

$$\left|\frac{\Delta\alpha}{\Delta\theta}\right| \approx \frac{4GM_{lens}D_{LS}}{c^2 R^2} , \quad \text{for } b < R \tag{4}$$

Therefore the amplification is constant for $b < R$. For $b > R$, it varies as $M_{lens}/b^2$, just as for a point mass.

In order to model a thick spherical lens, a series of 7 disk lenses was used. These were equally spaced in $z$ between $z = 0.5$ and $z = 2.0$, and each had 1/7$^{th}$ of the total mass. For each source point, a ray was stepped through those lenses toward $z = 0$, using the appropriate angular diameter distance for each lens. The amplifications are then added. The model predictions are shown as the dashed curves on the left plots in Fig. 3. The model parameters are $M_{lens} = 10^{21} M_\odot$, a lens radius of 350 Mpc, with the lens centered at approx. $(\alpha, \delta) = (195°, 0°)$. Its effects extend over a range of ±30° on the sky.

## 4. Discussion

The magnitude shift in the continuum spectra of the quasars toward $(\alpha, \delta) \sim (195°, 0°)$ is ~0.25 mag. The overall pattern of this apparent magnitude shift is <u>not</u> consistent with a Doppler shift due to our peculiar velocity toward that direction. The above discussion was for the *U* (ultraviolet) and *G* (green) filter bands. The *R, I,* and *Z* band spectra were found to have very similar shifts in filter magnitudes, both in size and their $(\alpha, \delta, z)$ dependence, while a Doppler shift would enhance the *U* band at the expense of the infrared. There is also no evidence for a redshift in the opposite hemisphere that would be characteristic of a large Doppler shift away from it. The angular pattern of the magnitude shifts cannot be explained as a systematic error in the filter magnitudes. All of the magnitude shifts are ~0.25 mag, while the quoted accuracies in the magnitudes are typically 0.03 in the *U* band for each quasar [2] and half that in the other bands. The quasar selection algorithms did not explicitly depend on angle, so the angular pattern cannot be explained by a variation in the completeness of the sample. Evidences from SDSS galaxy distributions and other data, which will be detailed in a longer article, are inconsistent with an explanation that attributes it to a systematic variation of the depth of the survey with angle. The hotspot is centered at a Galactic latitude of 63° and so it is well away from the Galactic disk.

The general features of the intensity enhancement seem to be reproduced quite well by the gravitational lensing of a bubble universe with mass $\sim 10^{21} M_\odot$, as described in Sect. 3. While this simple model cannot be expected to reproduce the details of the lensing, it does seem to exhibit the essential features quite well. If the bubble is expanding, time-dependent effects, such as the changing bubble radius, will distort the angular distribution from the simple model's expectation and change the *z* dependence. Since the $\Delta U$ are calculated by subtracting the average *U* at each *z* and the bubble includes such a large fraction of the data, the overall *z* dependence is partially washed out and the angular contrast is reduced. The error bars in Fig. 3 are calculated from the standard deviation of the $\Delta U$ for the <u>entire</u> quasar sample divided by the square root of the number of entries in the bin. This should be a generous estimate of the statistical error for each bin. The variations of the points are obviously not statistical and can be attributed to lensing on smaller scales by galaxies and clusters in the bubble and in the foreground. The angular distribution in Fig. 4 also shows evidence for partial Einstein arcs due to the strong gravitational lensing around the bubble. These appear near the upper left of the figure and near the center, and contribute to the large fluctuations in the magnitudes that are apparent in Fig. 3 left, especially near $\alpha = 230°$, for most of the *z* bins. The angular distribution in Fig. 4 is truncated below $\delta = -5°$ by the SDSS coverage, so the center of the hot spot probably lies at more negative $\delta$. A large gravitational lens seems to be the only way of explaining the systematic angular pattern of the enhancement that extends from $z \approx 0.5$ out to at least $z \approx 2.2$.

There are a number of other indications of new physics in this same angular range, which we can only summarize here. Abate and Feldman [6] see a similar enhancement in the photometric magnitudes of SDSS luminous red galaxies (LRG) for $z \sim 0.475$ over roughly the same angular region as the quasars, though their effect seems to decrease at $z \sim 0.625$. The maximum of this enhancement is located at roughly $\alpha = 180°$, with the data averaged over all SDSS declinations. The LRG enhancement is more than an order of magnitude larger than what would be expected from large-scale bulk flow measurements of peculiar velocities [7, 8, 9]. These bulk flow measurements are already $3\sigma$ higher than standard $\Lambda$CDM predictions. Thomas *et al.* [10] report a large excess of power in the statistical clustering of LRG in the photometric SDSS DR7 galaxy sample. This is seen for the same angular range as in Fig. 4 over the lowest multipoles in the angular power spectra, and it is most prominent in their highest redshift bin, $0.6 < z < 0.65$. [This "clustering" seems to be apparent in the number density distribution in Fig. 4.]

The direction of the quasar intensity enhancement is also close to that of the so-called "Axis of Evil", a name coined by K. Land and J. Magueijo [11] to describe the anomalies in the low multipoles of the CMB toward $(\alpha, \delta) \sim (173°, 4°)$. The extensive literature on the anomalies in the CMB was recently reviewed by Copi *et al.* [12].

Antoniou and Perivolaropoulos [13] also describe an asymmetry in the supernovae intensity distribution with the direction of maximum accelerating expansion rate at $\alpha \sim 200°$. They also review other evidence for a cosmological preferred axis. All of their anisotropy axes have right ascensions of approx. 180° with declinations ranging from –51° to 13°.

There has, of course, been considerable discussion of multiverses and bubble universes within or outside of our own universe. S. Feeney *et al.* [14] discuss possible observational tests to search for evidence of bubble universes. The "bulls eye" CMB temperature modulation they show in their Fig. 1, top left, looks strikingly like that in Fig. 4 for the quasar magnitude shifts except for the incidental color inversion.

It is remarkable that the presence of a massive bubble universe can explain not only the anomalies in the angular distribution of quasar intensities, but also the LRG and CMB anomalies, the bulk flow discrepancies, and possibly the supernovae magnitude enhancements, which are maximized along the same direction.

The confirmation of a large bubble universe will have profound implications for all of cosmology. Strict homogeneity and isotropy would no longer be the norm on the largest scales.

The bubble occupies a large portion of the sky near the north Galactic pole. Its lensing will distort the measured intensities of supernovae and other objects at large redshifts, and its gravitation will affect their apparent distances by shifting their spectral lines.

The dedicated efforts of the SDSS team and the support of their sponsors that made this work possible are gratefully acknowledged.